\documentclass[preprint,showpacs]{revtex4}
\usepackage{epsfig}
\usepackage{amsmath}
\usepackage{amssymb}

\begin{document}

\title{Branching ratio and CP violation of $B_{s}\to \pi K$  decays\\
in the perturbative QCD approach}

\author{Xian-Qiao Yu\footnote{yuxq@mail.ihep.ac.cn},
 Ying Li\footnote{liying@mail.ihep.ac.cn}}

\affiliation{
 {\it \small Institute of High
Energy Physics, P.O.Box 918(4), Beijing 100049, China;}\\
{\it \small  Graduate School of the Chinese Academy of
Sciences, Beijing 100049, China}}
\author{Cai-Dian L\"u\footnote{lucd@mail.ihep.ac.cn}}

\affiliation{{\it \small  CCAST (World Laboratory), P.O. Box
8730, Beijing 100080, China;}\\
 {\it \small  Institute of High
Energy Physics, P.O.Box 918(4), Beijing 100049,
China\footnote{Mailing address} }}

\date{\today}

\begin{abstract}
  In the framework of perturbative QCD approach, we calculate the
branching ratio and CP asymmetry for $B_{s}^0(\bar{B}_{s}) \to
\pi^{\pm} K^\mp$ and $B_{s}(\bar{B}_{s})\rightarrow
\pi^{0}\bar{K}^{0}(K^{0})$ decays. Besides the usual factorizable
diagrams, both non-factorizable and annihilation type
contributions are taken into account. We find that (a) the
branching ratio of $B_{s}^0(\bar{B}_{s}) \to \pi^{\pm} K^\mp$ is
about $(6-10) \times 10^{-6}$; $Br(B_{s}(\bar{B}_{s})\rightarrow
\pi^{0}\bar{K}^{0}(K^{0}))$ about $ (1-3) \times 10^{-7}$; and (b)
there are large CP asymmetries in the two processes, which can be
tested in the near future LHC-b experiments at CERN and BTeV
experiments at Fermilab.
\end{abstract}

 \maketitle

\section{Introduction}

  The rare charmless B meson decays  arouse  more and more interest,
since it is a good place for testing the Standard Model (SM),
studying CP violation and looking for possible new physics beyond
the SM. Since 1999, the B factories in KEK and SLAC collect more
and more data sample of rare B decays. In the future CERN Large
Hadron Collider beauty experiments (LHC-b), the heavier $B_s$ and
$B_c$ mesons can also be produced. With the bright hope in LHC-b
experiments and BTeV experiments at Fermilab, following a previous
study of $B_s\rightarrow \pi^{+}\pi^{-}$ decay \cite{1}, we
continue to investigate other $B_s$ rare decays.

 The most difficult problem in theoretical calculation of
 non-leptonic $B$ decays is the calculation of hadronic matrix
 element. The widely used method is the factorization
 approach (FA) \cite{2}. It is a great success in explaining the
 branching ratio of many decays \cite{3,5}, although it
 is a very simple method. In order to improve the theoretical
 precision, QCD factorization \cite{6} and perturbative QCD approach (PQCD)
 \cite{7}
 are developed. Perturbative QCD factorization theorem
 for exclusive heavy-meson decays has been proved some time ago,
 and applied to semi-leptonic $B \to D(\pi) l\nu$  decays \cite{7},
 the non-leptonic
 $B \to K \pi$ \cite{15}, $\pi\pi$ \cite{8} decays. PQCD is a method to
 factorize hard components from a
 QCD process, which can be treated by perturbation theory. Non-perturbative
 parts are organized in the form of universal hadron light cone wave functions, which
 can be extracted from experiments or constrained  by lattice calculations
 and QCD sum rules. More information about PQCD approach can be found in
\cite{7,pqcd}.

In this paper, we would like to study the $B_{s}^0(\bar{B}_{s})
\to \pi^{\pm} K^\mp$ and $B_{s}(\bar{B}_{s})\rightarrow
\pi^{0}\bar{K}^{0}(K^{0})$ decays in the perturbative QCD
approach. In our calculation, we ignore
 the soft final state interaction  because there are not many resonances
 near the energy region of $B_s$ mass. Our theoretical formulas for the decay $B_{s}\to\pi K$ in PQCD
framework are given in the next section. In section
\ref{sc:neval}, we give the numerical results of the branching
ratio of $B_{s}\to\pi K$ and discussions for CP asymmetries and
the form factor of $B_s \to K$ etc. At last, we give a short
summary in section \ref{summ}.

\begin{figure}[htb]
\vspace{0.5cm}
\begin{center}
\includegraphics[scale=0.75]{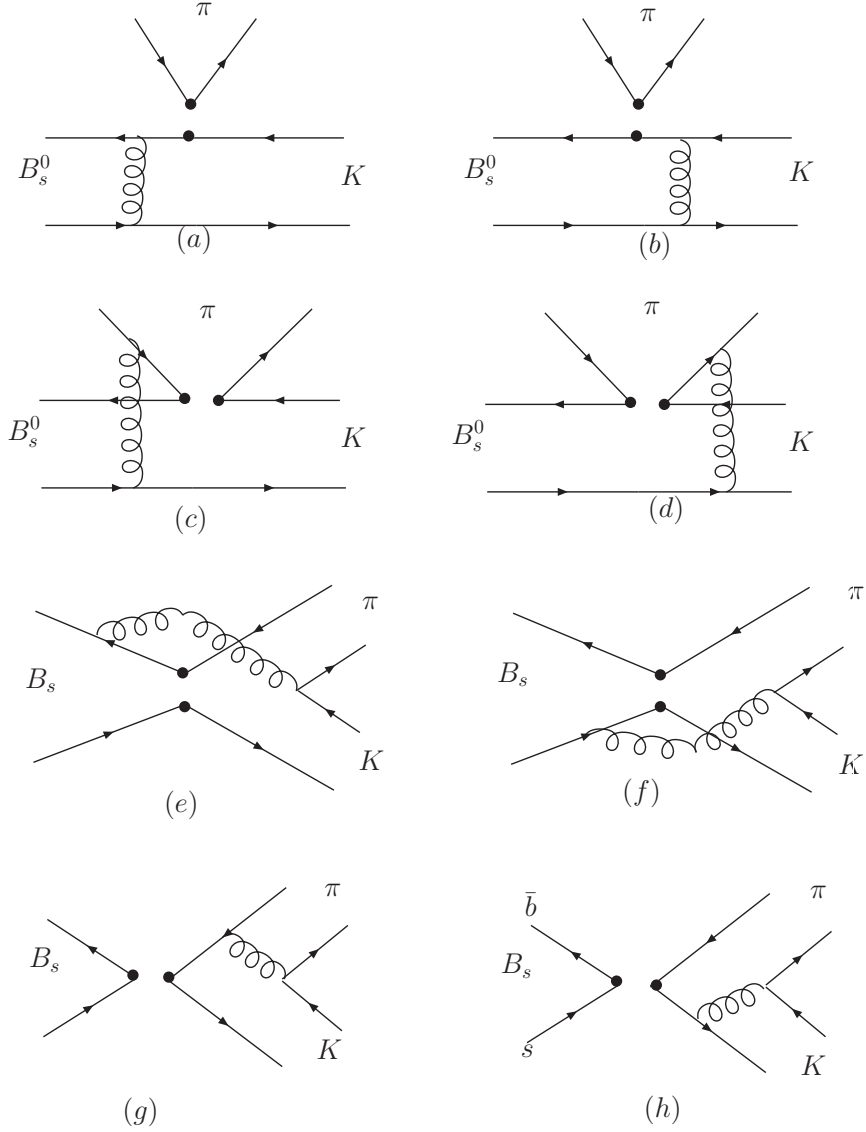}
\caption{The lowest order diagrams for $B_{s}^0 \to \pi K$ decay.}
\label{figure:Fig1}
\end{center}
\end{figure}

\section{Perturbative calculations}\label{sc:fm}

For decay $B_{s}\to \pi K$, the related effective Hamiltonian is
given by  \cite{9}
\begin{equation}
 H_\mathrm{eff} = \frac{G_F}{\sqrt{2}}\left\{ V_{ud}V_{ub}^* \left[
C_1(\mu) O_1(\mu) + C_2(\mu) O_2(\mu)
\right]-V_{tb}^*V_{td}\sum_{i=3}^{10}C_i(\mu)
O_i(\mu)\right\},\label{hami}
\end{equation}
 where $C_{i}(\mu)(i=1,\cdots,10)$ are Wilson coefficients at the
  renormalization scale $\mu$ and $O_{i}(i=1,\cdots,10)$ are the four quark operators
\begin{equation}\begin{array}{ll}
  O_1 = (\bar{b}_iu_j)_{V-A}(\bar{u}_jd_i)_{V-A},  &
  O_2 = (\bar{b}_iu_i)_{V-A} (\bar{u}_jd_j)_{V-A},  \\
  O_3 = (\bar{b}_id_i)_{V-A}\sum_{q} (\bar{q}_jq_j)_{V-A},  &
  O_4 = (\bar{b}_id_j)_{V-A}\sum_{q} (\bar{q}_jq_i)_{V-A}, \\
  O_5 = (\bar{b}_id_i)_{V-A}\sum_{q} (\bar{q}_jq_j)_{V+A},  &
  O_6 = (\bar{b}_id_j)_{V-A} \sum_{q} (\bar{q}_jq_i)_{V+A}, \\
  O_7 = \frac{3}{2}(\bar{b}_id_i)_{V-A} \sum_{q}
   e_q(\bar{q}_jq_j)_{V+A},   &
   O_8 = \frac{3}{2}(\bar{b}_id_j)_{V-A}\sum_{q} e_q
  (\bar{q}_jq_i)_{V+A}, \\
  O_9 = \frac{3}{2}(\bar{b}_id_i)_{V-A}\sum_{q}
  e_q(\bar{q}_jq_j)_{V-A}, &
   O_{10} = \frac{3}{2}(\bar{b}_id_j)_{V-A}\sum_{q}
 e_q(\bar{q}_jq_i)_{V-A}. \label{eq:effectiv}
 \end{array}
\end{equation}
Here $i$ and $j$ are $SU(3)$ color indices; the sum over $q$ runs
over the quark fields that are active at the scale $\mu=O(m_{b})$,
i.e., $q\in \{u,d,s,c,b\}$. Operators $O_{1}, O_{2}$ come from
tree level interaction, while $O_{3}, O_{4}, O_{5}, O_{6}$ are
QCD-Penguins operators and $O_{7}, O_{8}, O_{9}, O_{10}$ come from
electroweak-penguins.

Working at the rest frame of $B_s$ meson, we take kaon and pion
masses $M_{K}\sim M_{\pi}\sim 0$, which are much smaller than
$M_{B_s}$. In the light-cone coordinates, the momenta of the
$B_s$, $K$ and $\pi$ can be written as :
\begin{equation}
       P_1 = \frac{M_B}{\sqrt{2}} (1,1,{\bf 0}_T),\ \ \  P_2 =
       \frac{M_B}{\sqrt{2}} (0,1,{\bf 0}_T), \ \ \ P_3 =
       \frac{M_B}{\sqrt{2}} (1,0,{\bf 0}_T) . \label{eq:momentun1}
\end{equation}
Denoting the light (anti-)quark momenta in $B$, $K$ and $\pi$ as
$k_1$, $k_2$ and $k_3$, respectively, we can choose:
\begin{equation}
k_1 = (x_1p_1^+, 0, {\bf k}_{1T}), ~~~k_2 = (0, x_2p_2^-, {\bf
k}_{2T}),~~~
 k_3 = (x_3p_3^+ , 0, {\bf k}_{3T}). \label{eq:momentun2}
\end{equation}

In the following, we start to compute the decay amplitudes of
$B_{s}\rightarrow\pi K$.

According to effective Hamiltonian (\ref{hami}), we draw the
lowest order diagrams of $B_{s}\rightarrow\pi K$ in Fig.
\ref{figure:Fig1}. Let us first look at the usual factorizable
diagrams (a) and (b). they can give the $B_s\to K$ form factor if
take away the Wilson coefficients. The operators $O_1, O_2, O_3,
O_4, O_9$ and $O_{10}$ are $(V-A)(V-A)$ currents, and the sum of
their contributions is given by
\begin{multline}
F_e[C]  = 16\pi C_F M_B^2 \int_0^1\!\!\!
 dx_1 dx_2
  \int_0^\infty\!\!\!\!\! b_1 db_1\, b_2 db_2\
 \phi_B(x_1,b_1) \\
 \times \bigl\{[
 (2-x_{2} )\phi_{K}^A(x_2)-r_K(1-2x_{2})\phi_{K}^P(x_2)\\
 +r_K(1-2x_{2})\phi_{K}^T(x_2)]
 \alpha_{s}(t_{a}^{1})h_{a}(x_{1},1-x_2,b_1,b_2)
 \exp[-S_B(t_{a}^{1})-S_K(t_{a}^{1})]C(t_{a}^{1})\\
+2r_K\phi_{K}^P(x_2)\alpha_{s}(t_{a}^{2})
h_{a}(1-x_{2},x_1,b_2,b_1)\exp[-S_B(t_{a}^{2})-S_K(t_{a}^{2})]C(t_{a}^{2})
\bigr\},\label{fe}
\end{multline}
where $r_{\pi}=m_{0\pi}/m_B=m_{\pi}^2/[m_B(m_u+m_d)]$,
 $r_K=m_{0K}/m_B=m_{K}^2/[m_B(m_s+m_u)]$. $C_F=4/3$ is the
 group factor of the $SU(3)_{c}$ gauge group.
 The expressions of the meson distribution amplitudes
 $\phi_{M}$, the Sudakov factor $S_{X}(t_{i})(X=B_s, K,\pi)$,
 and the functions $h_a$ are given in the appendix. In above formula,
 the Wilson coefficients $C(t)$ of the corresponding operators
 are process dependent.

 The operator $O_5, O_6, O_7, O_8$ have the structure of $(V-A)(V+A)$,
  their amplitude is
\begin{multline}
F_e^{P}[C]  =32\pi C_F M_B^2 r_{\pi}\int_0^1\!\!\!
 dx_1 dx_2
  \int_0^\infty\!\!\!\!\! b_1 db_1\, b_2 db_2\
 \phi_B(x_1,b_1) \\
 \times \bigl\{[
 \phi_{K}^A(x_2)-r_K(x_{2}-3)\phi_{K}^P(x_2)\\
 +r_K(1-x_{2} )\phi_{K}^T(x_2)]
 \alpha_{s}(t_{a}^{1})h_{a}(x_{1},1-x_2,b_1,b_2)\exp[-S_B(t_{a}^{1})
 -S_K(t_{a}^{1})]C(t_{a}^{1})\\
+2r_K\phi_{K}^P(x_2)
\alpha_{s}(t_{a}^{2})h_{a}(1-x_{2},x_1,b_2,b_1)\exp[-S_B(t_{a}^{2})
-S_K(t_{a}^{2})]C(t_{a}^{2})\bigr\}.\label{fep}
\end{multline}

For the non-factorizable diagrams (c) and (d), all three meson
wave functions are involved. Using $\delta$  function
$\delta(b_1-b_3)$, the integration of $b_1$ can be preformed
easily. For the $(V-A)(V-A)$ operators the result is:
\begin{multline}
M_e[C]  =-\frac{32}{3}\pi C_F\sqrt{2N_{c}} M_B^2 \int_0^1\!\!\!
 dx_1 dx_2dx_3
  \int_0^\infty\!\!\!\!\! b_2 db_2\ b_3 db_3\
 \phi_B(x_1,b_3) \\
 \times \bigl\{[
 (x_3-1)\phi_{\pi}^A(x_3)\phi_{K}^A(x_2)+r_K(1-x_{2})\phi_{\pi}^A(x_3)
 \phi_{K}^P(x_2)
  +r_K(1-x_{2})\phi_{\pi}^A(x_3)\phi_{K}^T(x_2)]C(t_c^1)\\
 \alpha_{s}(t_{c}^{1})h_{c}^{(1)}(x_{1},x_2,x_3,b_2,b_3)\exp[-S_B(t_{c}^{1})
 -S_{\pi}(t_{c}^{1})-S_K(t_{c}^{1})]
  -[(x_2-x_3-1)\phi_{\pi}^A(x_3)\phi_{K}^A(x_2)\\
  +r_K(1-x_{2})\phi_{\pi}^A(x_3)\phi_{K}^P(x_2)
  -r_K(1-x_{2})\phi_{\pi}^A(x_3)\phi_{K}^T(x_2)]C(t_c^2)\\
\alpha_{s}(t_{c}^{2})h_{c}^{(2)}(x_{1},x_2,x_3,b_2,b_3)\exp[-S_B(t_{c}^{2})
-S_{\pi}(t_{c}^{2})-S_K(t_{c}^{2})]
 \bigr\}.\label{me}
\end{multline}
For the$(V-A)(V+A)$ operators, the formula is:
 \begin{multline}
M_e^{P}[C]  =-\frac{32}{3}\pi C_F\sqrt{2N_{c}} M_B^2
r_{\pi}\int_0^1\!\!\!
 dx_1 dx_2dx_3
  \int_0^\infty\!\!\!\!\! b_2 db_2\ b_3 db_3\
 \phi_B(x_1,b_3) \\
 \times
 \bigl\{[r_K(x_{2}+x_3-2)\phi_{\pi}^P(x_3)\phi_{K}^P(x_2)-r_K
 (x_2-x_3)\phi_{\pi}^P(x_3)\phi_{K}^T(x_2)-r_K(x_2-x_3)\phi_{\pi}^T(x_3)
 \phi_{K}^P(x_2)\\
  -r_K(2-x_{2}-x_3)\phi_{\pi}^T(x_3)\phi_{K}^T(x_2)-
    (1-x_3)\phi_{\pi}^P(x_3)\phi_{K}^A(x_2)-(1-x_3)\phi_{\pi}^T(x_3)
    \phi_{K}^A(x_2)]C(t_c^1)\\
 \alpha_{s}(t_{c}^{1})h_{c}^{(1)}(x_{1},x_2,x_3,b_2,b_3)\exp[-S_B(t_{c}^{1})
 -S_{\pi}(t_{c}^{1})-S_K(t_{c}^{1})]
  +[r_K(1-x_{2}+x_3)\phi_{\pi}^P(x_3)\phi_{K}^P(x_2)\\
  +r_K(x_2+x_3-1)\phi_{\pi}^P(x_3)\phi_{K}^T(x_2)
  -r_K(x_2+x_3-1)\phi_{\pi}^T(x_3)\phi_{K}^P(x_2)
  -r_K(1-x_{2}+x_3)\phi_{\pi}^T(x_3)\phi_{K}^T(x_2)\\
  +x_3\phi_{\pi}^P(x_3)\phi_{K}^A(x_2)-x_3\phi_{\pi}^T(x_3)\phi_{K}^A(x_2)]
  C(t_c^2)\\
\alpha_{s}(t_{c}^{2})h_{c}^{(2)}(x_{1},x_2,x_3,b_2,b_3)\exp[-S_B(t_{c}^{2})
-S_{\pi}(t_{c}^{2})-S_K(t_{c}^{2})]
 \bigr\}.\label{mep}
\end{multline}
Similar to (c),(d), the annihilation diagrams (e) and (f) also
involve all three meson wave functions. Here we have two kinds of
amplitudes, $M_a$ is the contribution containing the operator of
type $(V-A)(V-A)$, and $M_a^P$ is the contribution containing the
operator of type $(V-A)(V+A)$.
\begin{multline}
M_a[C]  =-\frac{32}{3}\pi C_F\sqrt{2N_{c}} M_B^2 \int_0^1\!\!\!
 dx_1 dx_2dx_3
  \int_0^\infty\!\!\!\!\! b_1 db_1\ b_2 db_2\
 \phi_B(x_1,b_1) \\
 \times
 \bigl\{[x_3\phi_{\pi}^A(x_3)\phi_{K}^A(x_2)+r_{\pi}r_K
 (2+x_2+x_3)\phi_{\pi}^P(x_3)\phi_{K}^P(x_2)-r_{\pi}r_K(x_2-x_3)\phi_{\pi}^P(x_3)
 \phi_{K}^T(x_2)\\
  -r_{\pi}r_K(x_2-x_3)\phi_{\pi}^T(x_3)\phi_{K}^P(x_2)-r_{\pi}r_K
    (2-x_2-x_3)\phi_{\pi}^T(x_3)\phi_{K}^T(x_2)]C(t_e^1)\\
 \alpha_{s}(t_{e}^{1})h_{e}^{(1)}(x_{1},x_2,x_3,b_1,b_2)\exp[-S_B(t_{e}^{1})
 -S_{\pi}(t_{e}^{1})-S_K(t_{e}^{1})]
  -[x_2\phi_{\pi}^A(x_3)\phi_{K}^A(x_2)\\
  +r_{\pi}r_K(x_2+x_3)\phi_{\pi}^P(x_3)\phi_{K}^P(x_2)
  +r_{\pi}r_K(x_2-x_3)\phi_{\pi}^P(x_3)\phi_{K}^T(x_2)
  +r_{\pi}r_K(x_2-x_3)\phi_{\pi}^T(x_3)\phi_{K}^P(x_2)\\
  +r_{\pi}r_K(x_2+x_3)\phi_{\pi}^T(x_3)\phi_{K}^T(x_2)]C(t_e^2)\\
\alpha_{s}(t_{e}^{2})h_{e}^{(2)}(x_{1},x_2,x_3,b_1,b_2)\exp[-S_B(t_{e}^{2})
-S_{\pi}(t_{e}^{2})-S_K(t_{e}^{2})]
 \bigr\},\label{ma}
\end{multline}
\begin{multline}
M_a^P[C]  =-\frac{32}{3}\pi C_F\sqrt{2N_{c}} M_B^2 \int_0^1\!\!\!
 dx_1 dx_2dx_3
  \int_0^\infty\!\!\!\!\! b_1 db_1\ b_2 db_2\
 \phi_B(x_1,b_1) \\
 \times
 \bigl\{[r_K(2-x_2)\phi_{\pi}^A(x_3)\phi_{K}^P(x_2)+r_K
 (2-x_2)\phi_{\pi}^A(x_3)\phi_{K}^T(x_2)-r_{\pi}(2-x_3)\phi_{\pi}^P(x_3)
 \phi_{K}^A(x_2)\\
  -r_{\pi}(2-x_3)\phi_{\pi}^T(x_3)\phi_{K}^A(x_2)]C(t_e^1)\\
 \alpha_{s}(t_{e}^{1})h_{e}^{(1)}(x_{1},x_2,x_3,b_1,b_2)\exp[-S_B(t_{e}^{1})
 -S_{\pi}
 (t_{e}^{1})-S_K(t_{e}^{1})]
  +[r_Kx_2\phi_{\pi}^A(x_3)\phi_{K}^P(x_2)\\
  +r_Kx_2\phi_{\pi}^A(x_3)\phi_{K}^T(x_2)
  -r_{\pi}x_3\phi_{\pi}^P(x_3)\phi_{K}^A(x_2)
  -r_{\pi}x_3\phi_{\pi}^T(x_3)\phi_{K}^A(x_2)]C(t_e^2)\\
\alpha_{s}(t_{e}^{2})h_{e}^{(2)}(x_{1},x_2,x_3,b_1,b_2)\exp[-S_B(t_{e}^{2})
-S_{\pi}(t_{e}^{2})-S_K(t_{e}^{2})]
 \bigr\}.\label{map}
\end{multline}

The factorizable annihilation diagrams (g) and (h) involve only
two light mesons wave functions. $F_a$ is for $(V-A)(V-A)$ type
operators, and $F_a^P$ is for $(V-A)(V+A)$ type operators:
\begin{multline}
F_a[C]  =16\pi C_F M_B^2 \int_0^1\!\!\!
 dx_2 dx_3
  \int_0^\infty\!\!\!\!\! b_2 db_2\, b_3 db_3\ \\
 \times \bigl\{[
 -x_{2}\phi_{\pi}^A(x_3)\phi_{K}^A(x_2)-2r_{\pi}r_K(1+x_{2})\phi_{\pi}^P(x_3)
 \phi_{K}^P(x_2)+
 2r_{\pi}r_K(1-x_{2})\phi_{\pi}^P(x_3)\phi_{K}^T(x_2)]\\
 \alpha_{s}(t_{g}^{1})h_{g}(x_2,x_3,b_2,b_3)\exp[-S_{\pi}(t_{g}^{1})-S_K(t_{g}^{1})]
 C(t_g^1)\\
+[x_3\phi_{\pi}^A(x_3)\phi_{K}^A(x_2)+2r_{\pi}r_K(1+x_3)\phi_{\pi}^P(x_3)
\phi_{K}^P(x_2)
 -2r_{\pi}r_K(1-x_3)\phi_{\pi}^T(x_3)\phi_{K}^P(x_2)]\\
 C(t_g^2)\alpha_{s}(t_{g}^{2})h_{g}(x_{3},x_2,b_3,b_2)
 \exp[-S_{\pi}(t_{g}^{2})-S_K(t_{g}^{2})]\bigr\},\label{fa}
\end{multline}
\begin{multline}
F_a^P[C] =32\pi C_F M_B^2 \int_0^1\!\!\!
 dx_2 dx_3
  \int_0^\infty\!\!\!\!\! b_2 db_2\, b_3 db_3\ \\
 \times \bigl\{[
  r_Kx_{2}\phi_{\pi}^A(x_3)\phi_{K}^P(x_2)-r_Kx_{2}\phi_{\pi}^A(x_3)
  \phi_{K}^T(x_2)+
 2r_{\pi}\phi_{\pi}^P(x_3)\phi_{K}^A(x_2)]\\
 \alpha_{s}(t_{g}^{1})h_{g}(x_2,x_3,b_2,b_3)\exp[-S_{\pi}(t_{g}^{1})-S_K(t_{g}^{1})]
 C(t_g^1)\\
+[2r_K\phi_{\pi}^A(x_3)\phi_{K}^P(x_2)+r_{\pi}x_3\phi_{\pi}^P(x_3)\phi_{K}^A(x_2)
 -r_{\pi}x_3\phi_{\pi}^T(x_3)\phi_{K}^A(x_2)]\\
C(t_g^2) \alpha_{s}(t_{g}^{2})h_{g}(x_{3},x_2,b_3,b_2)
 \exp[-S_{\pi}(t_{g}^{2})-S_K(t_{g}^{2})]\bigr\}.\label{fap}
\end{multline}

 From Equation (\ref{fe})-(\ref{fap}), the total decay amplitude
for $B_{s}\rightarrow\pi^{+}K^{-}$ can be written as
\begin{align}
 A(B_{s}^0\to\pi^{+}K^{-})\hspace*{11cm}\nonumber\\
=f_{\pi}F_{e}\left[V_{ud}V_{ub}^{*}(\frac{1}{3}C_{1}+C_{2})-V_{tb}^{*}V_{td}(
  \frac{1}{3}C_{3}+C_{4}+\frac{1}{3}C_{9}+C_{10})\right]\hspace*{3cm}\nonumber\\
  -f_{\pi}V_{tb}^{*}V_{td}F_{e}^{P}\left[\frac{1}{3}C_{5}+C_{6}
  +\frac{1}{3}C_{7}+C_{8}\right]
  +M_{e}\left[V_{ud}V_{ub}^{*}C_{1}-V_{tb}^{*}V_{td}(C_{3}+C_{9})\right]
  \hspace*{1cm}\nonumber\\
  -V_{tb}^{*}V_{td}M_e^{P}\left(C_{5}+C_{7}\right)-V_{tb}^{*}V_{td}M_a\left(C_{3}
  -\frac{1}{2}C_{9}\right)
  -V_{tb}^{*}V_{td}M_a^P\left(C_{5}-\frac{1}{2}C_{7}\right)\hspace*{2cm}\nonumber\\
  -f_{B} V_{tb}^{*}V_{td}F_a\left[\frac{1}{3}C_{3}+C_{4}-\frac{1}{6}C_{9}
  -\frac{1}{2}C_{10}\right]
   -f_{B}V_{tb}^{*}V_{td}F_a^{P}\left[\frac{1}{3}C_{5}+C_{6}-\frac{1}{6}C_{7}
   -\frac{1}{2}C_{8}\right], \label{eq:width}
\end{align}
and the decay width is expressed as
\begin{equation}
 \Gamma(B_s^0 \to \pi^+ K^-) = \frac{G_F^2 M_B^3}{128\pi}
|A(B_{s}^0\to\pi^{+}K^{-})|^2. \label{eq:width1}
\end{equation}
 The Wilson coefficient $C'_{i}s$ should be calculated at the
 appropriate scale t which can be found in the Appendix of Ref. \cite{8}.
 The decay amplitude of the charge conjugate channel $\bar{B}_{s}^{0}\to \pi^-
 K^+$ can be obtained by replacing $V_{ud}V_{ub}^{*}$ to
 $V_{ud}^{*}V_{ub}$ and $V_{tb}^{*}V_{td}$ to $V_{tb}V_{td}^{*}$
  in Eq.(\ref{eq:width}).

 For the decay $B_{s}\rightarrow\pi^{0}\bar{K}^{0}$, its amplitude
 can be written as
\begin{align}
 A(B_{s}^0\rightarrow\pi^{0}\bar{K}^{0})\hspace*{11cm}\nonumber\\
=f_{\pi}F_{e}\left[V_{ud}V_{ub}^{*}(C_{1}+\frac{1}{3}C_{2})-V_{tb}^{*}V_{td}(
  -\frac{1}{3}C_{3}-C_{4}+\frac{1}{6}C_{9}+\frac{1}{2}C_{10})\right]
  \hspace*{3.6cm}\nonumber\\
  -f_{\pi}V_{tb}^{*}V_{td}F_{e}^{p}\left[-\frac{1}{3}C_{5}-C_{6}+\frac{1}{6}C_{7}
  +\frac{1}{2}C_{8}\right]
  +M_{e}\left[V_{ud}V_{ub}^{*}C_{2}-V_{tb}^{*}V_{td}(-C_{3}+\frac{1}{2}C_{9})
  \right]\hspace*{1cm}\nonumber\\
  -V_{tb}^{*}V_{td}M_{e}^{p}\left(\frac{1}{2}C_{7}-C_{5}\right)-V_{tb}^{*}V_{td}M_{a}
  \left(\frac{1}{2}C_{9}-C_{3}\right)
  -V_{tb}^{*}V_{td}M_{a}^{p}\left(\frac{1}{2}C_{7}-C_{5}\right)\hspace*{2cm}
  \nonumber\\
  -f_{B}V_{tb}^{*}V_{td}F_{a}\left[-\frac{1}{3}C_{3}-C_{4}+\frac{1}{6}C_{9}
  +\frac{1}{2}C_{10}\right]
  -f_{B}V_{tb}^{*}V_{td}F_{a}^{p}\left[-\frac{1}{3}C_{5}-C_{6}
  +\frac{1}{6}C_{7}+\frac{1}{2}C_{8}\right].
\end{align}
 and the decay width is then expressed as
\begin{equation}
 \Gamma(B_s^0 \to \pi^0 \bar{K}^0) = \frac{G_F^2 M_B^3}{256\pi}
\left|A(B_{s}^0\rightarrow\pi^{0}\bar{K}^{0})\right|^2.
\label{eq:width2}
\end{equation}

\section{Numerical evaluation}\label{sc:neval}

The following parameters have been used in our numerical
calculation  \cite{13,nari}:
\begin{gather}
\nonumber  M_{B_s} = 5.37 \mbox{ GeV},  m_{0\pi} = 1.4 \mbox{
GeV}, m_{0K} = 1.6 \mbox{GeV}, \Lambda^{f=4}_{QCD}=0.25\mbox{
GeV}, f_{B_s} = 230 \mbox{ MeV},
\\ f_{\pi} = 130 \mbox{ MeV}, f_{K} = 160 \mbox{ MeV},
\tau_{B_s^0}=1.46\times 10^{-12}\mbox{s},
 | V_{tb}^{*}V_{td}|=0.0074, |V_{ub}^{*}V_{ud}|=0.0031.
\label{eq:shapewv}
\end{gather}
 We leave the CKM phase angle
$\alpha=\phi_2$ as a free parameter, whose definition is
\begin{equation}
 \alpha=\arg\Bigl[-\frac{V_{tb}^{*}V_{td}}{V_{ud}V_{ub}^{*}}
  \Bigr].
\end{equation}
 In this language, the decay amplitude of $B_{s}\rightarrow\pi^+
 K^-$ in eq.(\ref{eq:width})
can be parameterized as
 \begin{equation}
A = V_{ub}^*V_{ud}T-V_{tb}^*V_{td}P=V_{ub}^*V_{ud}T[1+z
e^{i(\alpha+\delta)}],\label{a1}
\end{equation}
where $z=|V_{tb}^*V_{td}/V_{ub}^*V_{ud}||P/T|$, and $\delta$ is
the relative strong phase between tree diagrams $T$ and penguin
diagrams $P$. $z$ and $\delta$ can be calculated from PQCD. Using
the above parameters in (\ref{eq:shapewv}), we get $z=22\%$ and
$\delta=134^{\circ}$  from PQCD calculation, which shows the
dominance of the tree contribution in this decay and a large
strong phase calculated from PQCD.

Similarly, the decay amplitude for
$\bar{B_{s}}\rightarrow\pi^{-}K^{+}$ can be parameterized as
\begin{equation}
 \bar{A}= V_{ub}V_{ud}^*T-V_{tb}V_{td}^*P=V_{ub}V_{ud}^*T[1+z
e^{i(-\alpha+\delta)}].\label{a2}
\end{equation}
 Therefore the averaged decay width for
$B_{s}^{0}(\bar{B}^{0}_{s})\rightarrow \pi^{\pm} K^{\mp}$ is
\begin{eqnarray}
 \Gamma(B_s^0(\bar{B}^{0}_{s}) \to \pi^\pm K^\mp) &=& \frac{G_F^2 M_B^3}{128\pi}
(|A|^2/2+|\bar{A}|^2/2)\hspace*{1cm}  \nonumber \\
&=&\frac{G_F^2 M_B^3}{128\pi}|V_{ub}^*V_{ud}T|^{2}[1+2z\cos\alpha
\cos\delta+z^{2}]. \label{eq:width3}
\end{eqnarray}
It is a function of $\cos\alpha \cos\delta$.

\begin{figure}[htbp]
  \begin{center}
  \epsfig{file=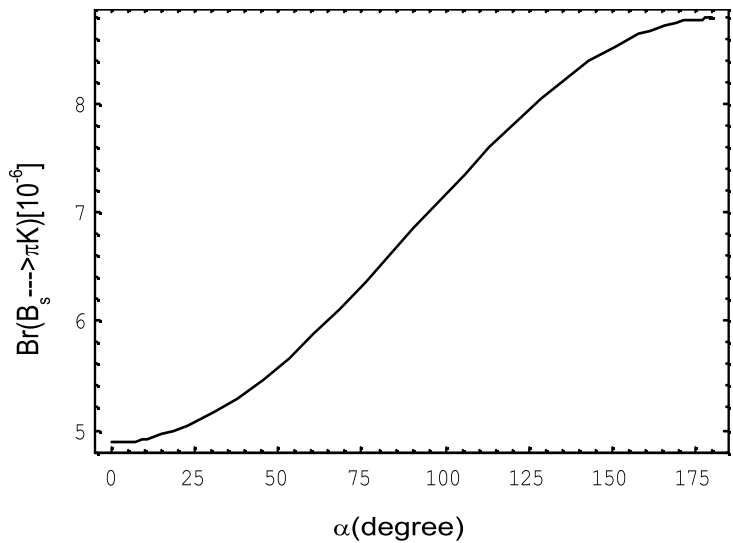,width=300pt,height=220pt}
   \end{center}
   \caption{The averaged branching ratio  of
   $B_{s}^0(\bar{B}_{s}) \to \pi^{\pm} K^{\mp}$ decay as a function of CKM angle
    $\alpha$.}
   \label{figure:Fig2}
  \end{figure}

 In Fig. \ref{figure:Fig2}, we plot the averaged branching ratio of the
decay $B_{s}^{0}(\bar{B}^{0}_{s})\rightarrow\pi^{\pm}K^{\mp}$ with
respect to the parameter $\alpha$. Since the latest experiment
constraint upon the CKM angle $\alpha$ from Belle and BaBar is
$\alpha$   around $100^{\circ}$  \cite{10},  we can arrive from
Fig. \ref{figure:Fig2}:
\begin{equation}
    6.2\times10^{-6}<Br(B_s^0(\bar{B}^{0}_{s}) \to \pi^\pm
    K^\mp)<8.1\times10^{-6},  \hspace{5mm}\text{for} 70^{\circ}<\alpha<130^{\circ}.
\end{equation}

Previous naive and generalized factorization approach gives a
similar branching ratios at $6-9\times 10^{-6}$ with the form
factor $F^{B_s \to K} \simeq 0.27$ \cite{fac}. In paper \cite{11},
Beneke {\it et.al} also calculate this decay mode using QCD
improved factorization approach (BBNS). It is based on naive
factorization approach. The dominant contribution is still
proportional to $B_s \to K$ form factor, which is introduced as an
input parameter. In principal, the decay amplitude expand as
series of $\alpha_s$ and $\Lambda/m_B$. But in practice, only the
first order of $\alpha_s$ corrections is calculated, including the
so called non-factorizable contributions. The annihilation type
contribution is power ($\Lambda/m_B$) suppressed in BBNS approach.
Therefore, the branching ratio predicted in QCD factorization and
PQCD should not differ too much; but the CP violation in these two
approaches will be different, since it depends on many non-leading
order contributions (See below for discussion). In Ref.\cite{11},
the branching ratio is about $10 \times 10^{-6}$, which is larger
than our PQCD result and previous FA method \cite{fac}, because
their form factor $F^{B_s\rightarrow K}(0)=0.31$ \cite{11} is
larger than the previous factorization approach and our
calculation below.

 The diagrams (a) and (b) in Fig.  \ref{figure:Fig1} correspond
  to the $B_{s}\rightarrow K$ transition form factor
  $F^{B_{s}\to K}(q^{2}=m_\pi^2\simeq 0)$, where $q=P_{1}-P_{2}$ is the momentum
   transfer. The sum of their amplitudes have been given by Eq.~(\ref{fe}),
   so we can use PQCD approach to compute this form factor.
  Our result is $F^{B_{s}\to K}(0)=0.27$, if $\omega_b=0.5$; and
  $F^{B_{s}\to K}(0)=0.32$, if $\omega_b=0.45$. In our approach,
   this form factor is sensitive to the decay constant and
   wave function of $B_s$ meson, where there is large uncertainty; but not
   sensitive to the $K$ meson wave function.
  Eventually this form factor can be extracted from
  semi-leptonic experiments $B_{s}\rightarrow K^{-}l^{+}\nu_{l}$ in the future.

 In our calculation, the only input parameters are wave
functions, which stand for the non-perturbative contributions. Up
to now, no exact solution is made for them. So the main
uncertainty in PQCD approach comes from $B_s, K,\pi$ wave
functions. In this paper, we choose the light cone wave functions
which are obtained from QCD Sum Rules \cite{atm,va}.  For $\pi$
meson, the distribution amplitude of light cone wave function
should take asymptotic form if the energy scale $\mu \to \infty$.
But in our case, the scale is not more than $5$GeV, so we
  choose the corrected asymptotic form for twist 2 distribution
 amplitude $\phi_{\pi}^A $, and other twist 3 distribution
 amplitudes
 derived using equation of motion by neglecting three particle
 wave functions \cite{va}. These
functions are listed in the Appendix, which  are also used in
decay mode $B \to K \pi $ \cite{15} and $B \to \pi \pi $ \cite{8}
etc.

We also try to use  the asymptotic form for $\pi$ meson, for all
the three distribution amplitudes $\phi_{\pi}^A $, $\phi_{\pi}^P $
and $ \phi_{\pi}^T $, since we have very poor knowledge about
twist 3 distribution amplitudes \cite{16}. The branching ratio of
$B_{s} \to \pi^{+} K^{-}$ is nearly unchanged (only $3\%$),
because the branching ratio of $B_{s}\to \pi^{+} K^{-}$ is mainly
determined by the form factor $F^{B_{s}\to K}(0)$ (see Fig.1(a)
and (b)) which is not dependent on $\pi$ wave function. However,
the CP asymmetry changes from $-28\%$ to $-13\%$ by $-54\%$, when
$\alpha =100^\circ$. This is because the direct CP asymmetry
depend on the strong phase (see discussion below), which comes
from non-factorizable and annihilation diagrams, where all three
meson wave functions are involved. The CP asymmetry predicted here
should be used with great care, since it depends on two much
uncertainties.

For heavy $B$ and $B_s$ meson, its wave function is still under
discussion using different approaches \cite{Bwave}. In this paper,
we find the branching ratio of $B_{s}^0(\bar{B}_{s}) \to \pi^{\pm}
K^\mp$ is sensitive to the wave function parameter $\omega_b$. For
$ 0.45< \omega_b <0.5 $, the resulted branching ratio will
decrease from about $10\times 10^{-6}$ to about $7\times 10^{-6}$.
When we set $\omega_b=0.45$, our result is more closer to that of
QCD factorization \cite{11}. This sensitive dependence should be
fixed by the $B_s \to K$ form factors from the semi-leptonic $B_s$
decays.  Other uncertainties in our calculation include the
next-to-leading order $\alpha_s$ QCD corrections and higher twist
contributions, which need more complicated calculations.

From our calculation, we find that the dominant contribution comes
from tree level diagrams (see Fig.1 (a) and
  (b)) in this decay. If SU(3) symmetry is good, the branching
  ratio of $B_{s}\to \pi^+ K^-$ should be   equal to that of $B^0\to
  \pi^{+}\pi^{-}$. The experimental result of $B^0\to
  \pi^{+}\pi^{-}$ is $Br(B\to
  \pi^{+}\pi^{-})=(4.3^{+1.6}_{-1.4}\pm0.5)\times 10^{-6}$
   \cite{12}. The predicted branching
  ratio of $B_{s}\to \pi K$ is about 1.7 times that of  $B_d\to
  \pi^{+}\pi^{-}$, where the difference  comes mainly from
     SU(3) symmetry breaking: the decay constant $f_{B_s}$ larger than $f_B$ and
     $f_K$ larger than $f_\pi$. In the calculation, we also find
     that the electroweak-penguins contribution is negligibly
     small as $0.001\%$ in branching ratio.

For the experimental side, there is recent upper limit on the
 decay $B_s^0 \to \pi^+ K^-$  \cite{18p},
\begin{equation}
    Br(B_s^0 \to \pi^+
    K^-)<7.5\times10^{-6},
\end{equation}
 at 90\% C.L. Our predicted result is consistent with this upper
 limit.

 \begin{figure}[tb]
  \begin{center}
  \epsfig{file=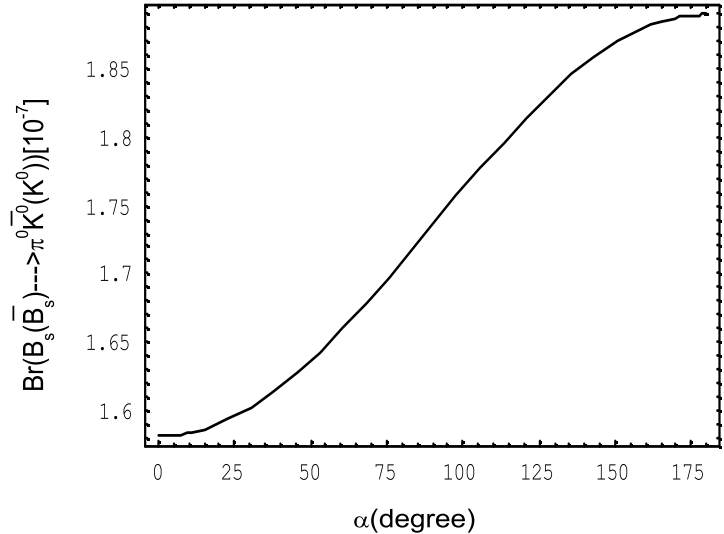,width=300pt,height=220pt}
   \end{center}
   \caption{The averaged branching ratio  of
   $B_{s}(\bar{B}_{s})\rightarrow \pi^{0}\bar{K}^{0}(K^{0})$ decay as a function
   of CKM angle $\alpha$.}
   \label{figure:Fig3}
  \end{figure}

  For the decays of $B_{s}(\bar{B}_{s})\rightarrow
  \pi^{0}\bar{K}^{0}(K^{0})$, the tree level contribution is
  suppressed due to the small Wilson coefficients $C_1+C_2/3$.
  Thus the penguin diagram contribution
  is comparable with the tree contribution. We study the averaged branching ratio of the
   decay $B_{s}(\bar{B}_{s})\rightarrow \pi^{0}\bar{K}^{0}(K^{0})$ as
   a function of $\alpha$ in Fig. \ref{figure:Fig3}. It is similar with
   Fig.\ref{figure:Fig2}. We find that the
   branching ratio of $B_{s}(\bar{B}_{s})\rightarrow \pi^{0}\bar{K}^{0}(K^{0})$
   is about $1.8\times10^{-7}$ when $\alpha$ is near $100^{\circ}$,
    it is a little smaller than the result of Ref.  \cite{11}.

\begin{figure}[tb]
  \begin{center}
  \epsfig{file=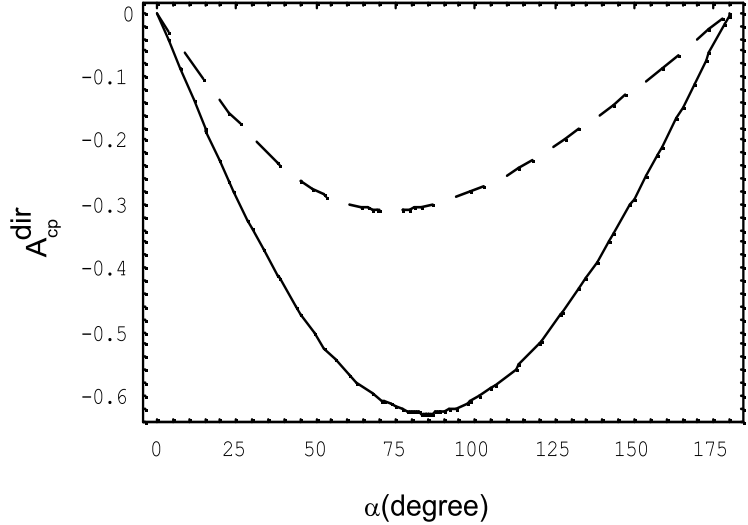,width=300pt,height=220pt}
   \end{center}
   \caption{Direct CP violation parameters of $B_{s}^0(\bar{B}_{s}) \to \pi^{\pm}
   K^{\mp}$ (dashed line) and $B_{s}(\bar{B}_{s})\rightarrow
   \pi^{0}\bar{K}^{0}(K^{0})$ (solid line) as
    a function of CKM angle $\alpha$.}
   \label{figure:Fig4}
  \end{figure}

  In SM, the CKM phase angle is the origin of CP violation. Using
  Eqs.(\ref{a1}) and (\ref{a2}), the direct CP
 violation parameter  can be derived as
 \begin{equation}
A_{CP}^{dir}=\frac{|A|^{2}-|\bar{A}|^{2}}{|A|^{2}+|\bar{A}|^{2}}
=\frac{-2z\sin\alpha\sin\delta}{1+2z\cos\alpha\cos\delta+z^{2}} .
\label{dcpv}
 \end{equation}
 It is approximately proportional to CKM angle $\sin \alpha$,
 strong phase $\sin \delta$ and the relative size $z$ between penguin
 contribution and tree contribution.
  We show the direct CP violation parameters as a function of CKM
 angle $\alpha$ in Fig. \ref{figure:Fig4}. From this figure one
 can see that the direct CP asymmetry parameter of $B_s^0(\bar{B}^{0}_{s})
 \to \pi^\pm
 K^\mp$ and $\pi^{0}\bar{K}^{0}(K^{0})$ can be as large as
 $-31\%$ and $-62\%$ when $\alpha$ is near $75^{\circ}$.
 The larger direct CP asymmetry of $B_s^0(\bar{B}^{0}_{s}) \to
 \pi^{0}\bar{K}^{0}(K^{0})$ decay is mainly due to a larger $z$ in
 $B_s^0(\bar{B}^{0}_{s}) \to
 \pi^{0}\bar{K}^{0}(K^{0})$ than in $B_s^0(\bar{B}^{0}_{s})
 \to \pi^\pm
 K^\mp$.

 The direct CP asymmetry predicted in QCD factorization approach
 is quite different from our result, due to the different source
 of strong phases. In QCD factorization approach, the strong phase
 mainly comes from the perturbative charm quark loop diagram,
 which is $\alpha_s$ suppressed \cite{11}. While the strong phase in PQCD
 comes mainly from non-factorizable and annihilation type
 diagrams. The sign of the direct CP asymmetry is different for
 these two approaches in $B_s^0(\bar{B}^{0}_{s})
 \to \pi^\pm K^\mp$ decay, and the magnitude of CP asymmetry in
 QCD factorization (about 5\%) is also smaller than PQCD.
  The   future LHC-b experiments can make a test for the two
  methods.

\begin{figure}[tb]
  \begin{center}
  \epsfig{file=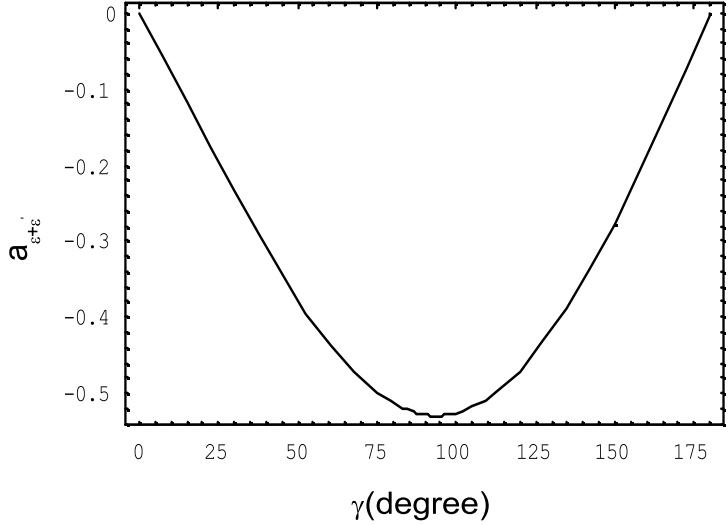,width=300pt,height=220pt}
   \end{center}
   \caption{Mixing induced CP violation parameter of   $B_{s}(\bar{B}_{s})\rightarrow
   \pi^{0}\bar{K}^{0}(K^{0})$   as
    a function of CKM angle $\gamma$.}
   \label{cp}
  \end{figure}

For the decays of $B_{s}(\bar{B}_{s})\rightarrow
  \pi^{0}\bar{K}^{0}(K^{0})$, the final $\bar{K}^{0}(K^{0})$
  mesons can not be detected directly. What the experiments
  measured are their mixtures $K_s$ and $K_L$, thus a mixing induced CP violation is involved. Following
  notations in the previous literature \cite{cp}, we define the mixing induced CP
  violation parameter as
 \begin{equation}
a_{\epsilon+\epsilon^{'}}=\frac{-2Im(\lambda_{CP})
}{1+|\lambda_{CP}|^{2}},\label{mcp}
 \end{equation}
 where
 \begin{equation}
\lambda_{CP}=\frac{V_{tb}^{*}V_{ts}\langle
\pi^{0}K^{0}|H_{eff}|\bar{B}_{s}^{0}\rangle}{V_{tb}V_{ts}^{*}\langle
\pi^{0}\bar{K}^{0}|H_{eff}|B_{s}^{0}\rangle}          .
 \end{equation}
Using unitarity condition of the CKM matrix $V_{tb}V_{td}^* = -
V_{ub} V_{ud}^* - V_{cb}V_{cd}^*$, and Eqs.(\ref{a1},\ref{a2}), we
can get
\begin{equation}
\lambda_{CP} = \frac{e^{-i\gamma} + x}{e^{i\gamma} +x},
\label{lam}
\end{equation}
where $x= \frac{V_{cb} V_{cd}^*}{|V_{ub}V_{ud}^*|}\frac{P}{T+P}$.
Combining eq.(\ref{lam}) and (\ref{mcp}), we can get
\begin{equation}
a_{\epsilon+\epsilon'} = \frac{\sin 2\gamma +2 Re(x) \sin \gamma
}{1+|x|^2 + 2 Re(x)\cos \gamma }. \label{cpf}
\end{equation}
If $|x|$ is a very small number, the mixing induced CP asymmetry
is proportional to $\sin 2\gamma$, which will be a good place for
the CKM angle $\gamma$ measurement. However as we already
mentioned, the tree contribution in this channel is suppressed,
$|x| = 2.3$ is a large number, so that the $\sin \gamma$ behavior
is dominant in the eq. (\ref{cpf}). The result of mixing induced
CP violation is shown in Fig. \ref{cp}, which is indeed a roughly
$\sin \gamma$ behavior. The tail near $\gamma \sim 180^\circ$ also
shows the contribution from $\sin 2\gamma$ in eq.(\ref{cpf}).

\section{Summary} \label{summ}

  In this work, we study the branching ratio and CP asymmetry of
  the decays $B_{s}^{0}(\bar{B}^{0}_{s})\rightarrow\pi^{\pm}K^{\mp}$ and
  $B_{s}(\bar{B}_{s})\rightarrow \pi^{0}\bar{K}^{0}(K^{0})$
  in PQCD approach.   From our calculation, we find that the branching
  ratio of $B_{s}^{0}(\bar{B}^{0}_{s})\rightarrow\pi^{\pm}K^{\mp}$
   is about $ (6\sim 10)\times10^{-6}$; $Br(B_{s}(\bar{B}_{s})\rightarrow
\pi^{0}\bar{K}^{0}(K^{0}))$ around  $2\times 10^{-7}$ and there
are large CP violation in the processes, which may be measured in
the
 future LHC-b experiments
  and BTeV experiments at Fermilab.

\section*{Acknowledgments}

 The authors thank M-Z Yang for helpful discussions, they also
 thank Professor Dong-Sheng Du for reading the manuscript.
  This work is partly supported by National Science Foundation of
 China under Grant No. 90103013, 10475085 and 10135060.


\begin{appendix} \label{appendix}

\section{formulas for the calculations used in the text}

 In the appendix we present the explicit expressions of the
 formulas used in section II. First, we give the expressions of the
 meson distribution amplitudes
 $\phi_{M}$. For $B_{s}$ meson wave function, we use the similar wave function as
  $B$ meson  \cite{8,15}:
\begin{equation}
\phi_{B_s}(x,b) = N_{B_s} x^2(1-x)^2 \exp \left[ -\frac{M_{B_s}^2\
x^2}{2 \omega_b^2} -\frac{1}{2} (\omega_b b)^2
\right].\label{waveb}
\end{equation}
We set the central value of parameter $\omega_b=0.5\mbox{ GeV}$ in
our numerical calculation, and $N_{B_s}=63.7\mbox{GeV}$ is the
normalization constant using $f_{B_s}=230 \mbox{MeV}$.

The $\pi$ meson's distribution amplitudes are given by light cone
QCD sum rules  \cite{va}:
\begin{eqnarray}
\nonumber \phi_{\pi}^A(x) &=& \frac{3f_{\pi}}{\sqrt{2 N_c}} x(1-x)
\left\{ 1 +0.44C_2^{3/2}(t) + 0.25C_4^{3/2}(t) \right\}, \\
\nonumber \phi_{\pi}^P(x) &=& \frac{f_{\pi}}{2\sqrt{2 N_c}}
\left\{ 1 +0.43C_2^{1/2}(t) +
0.09C_4^{1/2}(t) \right\}, \\
\phi_{\pi}^T(x) &=& \frac{f_{\pi}}{2\sqrt{2 N_c}}(1-2x) \left\{
1+0.55(10x^2-10x+1) \right\},
\end{eqnarray}
where $t=1-2x$. The Gegenbauer polynomials are defined by:
\begin{eqnarray}
C_2^{1/2}(t)=\frac{1}{2}(3t^2-1),\hspace{1cm} C_4^{1/2}(t)
=\frac{1}{8}(35t^4-30t^2+3),\nonumber\\
C_2^{3/2}(t)=\frac{3}{2}(5t^2-1),\hspace{1cm}
C_4^{3/2}(t)=\frac{15}{8}(21t^4-14t^2+1).
\end{eqnarray}

 We use the distribution amplitude $\phi^{A,P,T}_{K}$ of the K meson from
 Ref.  \cite{atm}:
 \begin{eqnarray}
\nonumber \phi_{K}^A(x) &=& \frac{6f_{K}}{2\sqrt{2
N_c}}x(1-x)[1+0.15t+0.405(5t^2-1)], \\
\nonumber \phi_{K}^P(x) &=& \frac{f_{K}}{2\sqrt{2 N_c}} [1
+0.106(3t^2-1)-0.148(3-30t^2+35t^4)/8], \\
\phi_{K}^T(x) &=& \frac{f_{K}}{2\sqrt{2 N_c}}t[1+0.1581(5t^2-3)],
\end{eqnarray}
whose coefficients correspond to $m_{0K} = 1.6 \mbox{GeV}$.

In our numerical analysis,  we use the one loop expression for the
strong running coupling constant,
\begin{equation}
 \alpha_s(\mu)=\frac{4\pi}{\beta_0\mathrm{ln}(\mu^2/\Lambda^2)},
\end{equation}
where $\beta_0=(33-2n_{f})/3$ and $n_{f}$ is the number of active
quark flavor at the appropriate scale $\mu$. $\Lambda$ is the QCD
scale, which we take $\Lambda=250 $MeV at $n_{f}=4$.

  $S_{B_s}$, $S_{\pi^{+}}$, $S_{k^{-}}$ used in the decay amplitudes
 are defined as
\begin{gather}
 S_{B_s}(t)=s(x_1P_1^+,b_1)+2\int_{1/b_1}^t\!\!\!\frac{d\bar\mu}{\bar\mu}
 \gamma(\alpha_s(\bar\mu)),\hspace*{3.5cm}\\
 S_{\pi^+}(t)=s(x_3P_3^+,b_3)+s((1-x_3)P_3^+,b_3)+
 2\int_{1/b_3}^t\!\!\!\frac{d\bar\mu}{\bar\mu}\gamma(\alpha_s(\bar\mu)),\\
 S_{K^-}(t)=s(x_2P_2^-,b_2)+s((1-x_2)P_2^-,b_2)+
 2\int_{1/b_2}^t\!\!\!\frac{d\bar\mu}{\bar\mu}\gamma(\alpha_s(\bar\mu)),
\end{gather}
where the so called Sudakov factor $s(Q,b)$  resulting from the
resummation of double logarithms is given as \cite{19,20}
\begin{equation}
s(Q,b)=\int_{1/b}^Q\!\!\! \frac{d\mu}{\mu}\Bigl[
\ln\left(\frac{Q}{\mu}\right)A(\alpha(\bar\mu))+B(\alpha_s(\bar\mu))
\Bigr] \label{su1}
\end{equation}
with
\begin{gather}
A=C_F\frac{\alpha_s}{\pi}+\left[\frac{67}{9}-\frac{\pi^2}{3}-\frac{10}{27}n_{f}+
\frac{2}{3}\beta_0\ln\left(\frac{e^{\gamma_E}}{2}\right)\right]
 \left(\frac{\alpha_s}{\pi}\right)^2 ,\\
B=\frac{2}{3}\frac{\alpha_s}{\pi}\ln\left(\frac{e^{2\gamma_{E}-1}}{2}\right).\hspace{6cm}
\end{gather}
Here $\gamma_E=0.57722\cdots$ is the Euler constant, $n_{f}$ is
the active quark flavor number. For the detailed derivation of the
Sudakov factors, see Ref.  \cite{7,21}.

The functions $h_{i}(i=a,c.e.g)$   come from the Fourier
transformation of propagators of virtual quark and gluon in the
hard part calculations. They are given as
\begin{align}
& h_{a}(x_1,x_2,b_1,b_2) = S_{t}(x_2)K_{0}(M_{B}\sqrt{x_1x_2}b_{1})\nonumber \\
&
  \times \bigl[\theta(b_2-b_1)I_{0}(M_{B}\sqrt{x_{2}}b_1)K_{0}(M_{B}\sqrt{x_{2}}b_2)+(b_1
\leftrightarrow b_2)\bigr],
 \label{eq:propagator1}
\end{align}

\begin{align}
& h_{c}^{(j)}(x_1,x_2,x_3,b_2,b_3) = \nonumber \\
&
\biggl\{\theta(b_2-b_3)I_{0}(M_B\sqrt{x_1(1-x_2)}b_3)K_{0}(M_B\sqrt{x_1(1-x_2)}b_2)
\nonumber \\
& \qquad\qquad\qquad\qquad + (b_2\leftrightarrow b_3) \biggr\}
 \times\left(
\begin{matrix}
 \mathrm{K}_0(M_B F_{(j)} b_3), & \text{for}\quad F^2_{(j)}>0 \\
 \frac{\pi i}{2} \mathrm{H}_0^{(1)}(M_B\sqrt{|F^2_{(j)}|}\ b_3), &
 \text{for}\quad F^2_{(j)}<0
\end{matrix}\right),
\label{eq:propagator2}
\end{align}
where $\mathrm{H}_0^{(1)}(z) = \mathrm{J}_0(z) + i\,
\mathrm{Y}_0(z)$, and $F_{(j)}$'s are defined by
\begin{equation}
 F^2_{(1)} = x_1+x_2+x_3-x_1x_2-x_2x_3-1,\
F^2_{(2)} = x_1-x_3-x_1x_2+x_2x_3;
\end{equation}
\begin{align}
& h^{(j)}_e(x_1,x_2,x_3,b_1,b_2) = \nonumber \\
& \biggl\{\theta(b_2-b_1) \frac{\pi i}{2}
\mathrm{H}_0^{(1)}(M_B\sqrt{x_2x_3}\, b_2)
 \mathrm{J}_0(M_B\sqrt{x_2x_3}\, b_1)
\nonumber \\
& \qquad\qquad\qquad\qquad + (b_1 \leftrightarrow b_2) \biggr\}
 \times\left(
\begin{matrix}
 \mathrm{K}_0(M_B F_{e(j)} b_1), & \text{for}\quad F^2_{e(j)}>0 \\
 \frac{\pi i}{2} \mathrm{H}_0^{(1)}(M_B\sqrt{|F^2_{e(j)}|}\ b_1), &
 \text{for}\quad F^2_{e(j)}<0
\end{matrix}\right),
\label{eq:propagator3}
\end{align}
where $F_{e(j)}$'s are defined by
\begin{equation}
 F^2_{e(1)} =x_1 +x_2+x_3-x_1x_2-x_2x_3,\
F^2_{e(2)} = x_1x_2-x_2x_3;
\end{equation}
  \begin{align}
& h_{g}(x_2,x_3,b_2,b_3) = S_{t}(x_2)\frac{\pi i}{2}H_{0}^{(1)}
(M_{B}\sqrt{x_2x_3}b_{3})\nonumber \\
&\times \bigl[\theta(b_3-b_2)J_{0}(M_{B}\sqrt{x_2}b_2)\frac{\pi
i}{2}
  H_{0}^{(1)}(M_{B}\sqrt{x_2}b_3)+(b_2
\leftrightarrow b_3)\bigr].
 \label{eq:propagator4}
\end{align}

 We adopt the parametrization for $S_{t}(x)$ contributing to the
 factorizable diagrams  \cite{22},
 \begin{align}
S_{t}(x)=\frac{2^{1+2c}\Gamma(3/2+c)}{\sqrt{\pi}\Gamma(1+c)}[x(1-x)]^{c},
\hspace{0.5cm}c=0.3.
\end{align}
The hard scale $t_{i}'s$ in Eq.(\ref{fe})-(\ref{fap}) are chosen
as
 \begin{eqnarray}
\nonumber t_a^{1} &=& \mathrm{max}(M_B \sqrt{1-x_{2}},
 1/b_1,1/b_2), \\
\nonumber t_a^{2} &=& \mathrm{max}(M_B \sqrt{x_{1}},
 1/b_1,1/b_2), \\
\nonumber t_c^{1} &=& \mathrm{max}(M_B
\sqrt{|F^2_{(1)}|},M_B\sqrt{x_{1}(1-x_2)},
 1/b_2,1/b_3), \\
 \nonumber t_c^{2} &=& \mathrm{max}(M_B
\sqrt{|F^2_{(2)}|},M_B\sqrt{x_{1}(1-x_2)},
 1/b_2,1/b_3), \\
 \nonumber t_e^{1} &=& \mathrm{max}(M_B
\sqrt{|F^2_{e(1)}|},M_B\sqrt{x_{2}x_3},
 1/b_1,1/b_2), \\
 \nonumber t_e^{2} &=& \mathrm{max}(M_B
\sqrt{|F^2_{e(2)}|},M_B\sqrt{x_{2}x_3},
 1/b_1,1/b_2), \\
 \nonumber t_g^{1} &=& \mathrm{max}(M_B \sqrt{x_{2}},
 1/b_2,1/b_3), \\
 t_g^{2} &=& \mathrm{max}(M_B \sqrt{x_{3}},
 1/b_2,1/b_3).
\end{eqnarray}
 They are given as the maximum energy scale appearing in each diagram
 to kill the large logarithmic radiative corrections.
\end{appendix}


\end{document}